\documentclass[12pt,a4paper]{article}
\pdfoutput=1

\usepackage[margin=1in]{geometry}
\usepackage[compress,numbers,sort]{natbib}
\usepackage{orcidlink}
\usepackage{graphicx}
\usepackage{wrapfig}
\usepackage{xcolor}
\usepackage{hyperref}
\usepackage{comment}
\usepackage{array}
\usepackage{tabularray}
\usepackage[T1]{fontenc}
\usepackage{amssymb,slashed,latexsym}
\usepackage{bm}
\usepackage{braket}
\usepackage{placeins}
\usepackage{adjustbox}
\usepackage{enumitem}
\usepackage{slashed}
\usepackage{color}
\usepackage{verbatim}
\usepackage{latexsym}
\usepackage{amsmath}
\usepackage{amsfonts}
\usepackage{graphicx}
\usepackage{bm}
\usepackage{adjustbox}
\usepackage{multirow}
\usepackage{mathrsfs}
\usepackage{adjustbox}
\usepackage{amssymb}
\usepackage{physics}
\usepackage{slashed}    
\usepackage{caption}   
\usepackage{subcaption} 
\usepackage[normalem]{ulem}
\usepackage{orcidlink}
\usepackage{mathtools}
\usepackage[capitalise]{cleveref}
\usepackage{csquotes}
\usepackage{siunitx}
\usepackage{comment}

\def\beq{\begin{equation}}
\def\eeq{\end{equation}}
\def\bea{\begin{eqnarray}}
\def\eea{\end{eqnarray}}

\newcolumntype{P}[1]{>{\centering\arraybackslash}p{#1}}
\newcolumntype{M}[1]{>{\centering\arraybackslash}m{#1}}
\UseTblrLibrary{booktabs}

\definecolor{lightblue}{rgb}{0.1, 0.5, 1.0}
\definecolor{darkblue}{cmyk}{1,0.4,0,0.3}
\definecolor{violet}{cmyk}{0,1,0,0.2}

\newcommand{\bld}[1]{\boldsymbol{#1}}

\hypersetup{
  colorlinks    = true, 
  urlcolor      = green,
  linkcolor     = blue, 
  citecolor     = blue,
  anchorcolor   = blue,
  menucolor     = blue,
  linktocpage   = true,
  pdfproducer   = medialab,
}

\crefname{section}{Sec.}{Sec.}

\begin{document}
\centerline{\Large\bf Higgs Criticality and the Metastability Bound:} 
\centerline{\Large\bf a target for future colliders} 
\vspace{1cm}
\vskip 12pt

\begin{center}
{\bf 
Maximilian Detering\orcidlink{0009-0001-1408-8192}$^{a}$
Victor Enguita\orcidlink{0000-0001-5977-9635}$^{b}$, 
Belen Gavela\orcidlink{0000-0002-2321-9190}$^{b}$, 
Thomas Steingasser\orcidlink{0000-0002-1726-2117}$^{c,d}$,
Tevong You\orcidlink{0000-0003-2391-7463}$^{a}$
}\\[7mm]

{\it $^a$Theoretical Particle Physics and Cosmology Group, Department of Physics,
King’s College London, London, WC2R 2LS, UK\\
\it $^b$Departamento de Fisica Teorica, Universidad Autonoma de Madrid, \\ 
and IFT-UAM/CSIC, Cantoblanco, 28049, Madrid, Spain}\\[1mm]
{\it $^c$Department of Physics, Massachusetts Institute of Technology, \\ Cambridge, MA 02139, USA}\\[1mm]
{{\it $^d$Black Hole Initiative at Harvard University, 20 Garden Street, \\ Cambridge, MA 02138, USA}}

\end{center}

\begin{abstract}
    New physics at the TeV scale or lower may destabilise the electroweak vacuum. How low could the vacuum instability scale be? This fundamental question may be tied to a deeper understanding of the Higgs potential and its associated hierarchy problem. The scale of vacuum instability can be viewed as an upper bound on the Higgs mass---the so-called vacuum metastability bound---and criticality of the Higgs potential through some underlying mechanism then places our universe at this metastable point. In this report, we summarise recent work developing this eminently testable hypothesis. If the vacuum metastability bound plays a role in determining the properties of the Higgs boson, the new physics responsible will likely be discovered or excluded in the entire natural region of parameter space at future facilities. This makes it a tantalising and attractive target for future colliders.
\end{abstract}

\thispagestyle{empty}
\setcounter{page}{0}

\newpage

\section{Motivation}

\subsection{The hierarchy problem as a fundamental challenge to modern particle physics}

In our modern understanding, quantum field theories are really Effective Field Theories (EFTs) whose Lagrangian terms are controlled by symmetries, thus leading to the expectation of ``natural'' values for their coefficients (see e.g.~Ref.~\cite{Craig:2022eqo} for a review). The limitations of this paradigm have been revealed by two of the most consequential open problems of modern fundamental physics, the electroweak (EW) hierarchy problem and the cosmological constant problem. In the Standard Model (SM), the cosmological constant and Higgs mass are unprotected by symmetry. Quantum fluctuations therefore contribute large corrections to their bare parameter values (which are expected to be calculable in terms of more fundamental parameters in the UV theory) up to the cut-off scale of the EFT that could be as high as the Planck scale. The tiny vacuum energy and Higgs mass measured in our universe must then be the result of fine-tuned cancellations to an inconceivable number of digits between the bare parameters and unrelated quantum corrections. In the case of the Higgs mass, such a contrived situation could be avoided if new physics with additional symmetries exist close to the EW scale. 

The absence of such new physics at the LHC and the failure of symmetry-based solutions for addressing the cosmological constant motivate exploring alternative solutions to the hierarchy problem. Starting with the so-called relaxion mechanism~\cite{Graham:2015cka}, the last decade has seen renewed speculations about the observed Higgs mass originating from a landscape of values with some cosmological selection mechanism. Here we summarise a different approach based on the observation that avoiding the instability and collapse of the electroweak vacuum places an upper bound on the Higgs mass. This so-called vacuum metastability bound could then explain the hierarchy problem if a dynamical mechanism for achieving criticality of the Higgs potential selects a metastable vacuum at the edge of a quantum phase transition. 

\subsection{Metastability bound}\label{sec:MSBound}
The increasing pressure experiments put on natural symmetry-based solutions to the EW hierarchy problem calls for the exploration of possibilities beyond the paradigm of symmetry. The hierarchy problem can be viewed as understanding the scale of the EW vacuum, determined by the Higgs potential,
\begin{equation}
    V(H) = - \mu_H^2 H^\dagger H + \lambda_H (H^\dagger H)^2 ,
\end{equation}
where $H$ is the Higgs doublet and the Higgs parameters receive radiative corrections. While the EW symmetry is broken spontaneously for $-\mu_H^2 < 0$, the existence of the EW vacuum also requires $\lambda_H > 0$. The renormalisation group (RG) evolution of the Higgs quartic coupling can lead to an instability of the Higgs potential with a negative quartic coupling in the ultraviolet (UV). A non-trivial vacuum may therefore only exist below the \textit{instability scale} $\mu_I$ given by the renormalisation scale of vanishing quartic coupling,
\begin{equation}\label{eq:instability-scale}
    \lambda (\mu_I) = 0,
\end{equation}
where $\lambda$ represents the quartic coupling including relevant loop-level corrections. 

\begin{wrapfigure}{l}{0.5\textwidth}
    \centering
    \includegraphics[width=\linewidth]{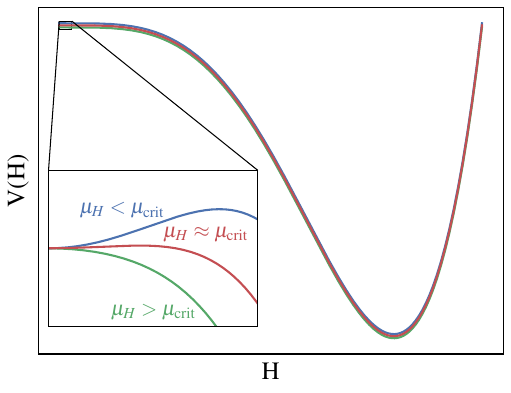}
    \caption{\it Shape of the Higgs potential for different values of the mass parameter $\mu_H^2$. The IR vacuum only exists for small values of the mass parameter.}
    \label{fig:metastability-bound}
\end{wrapfigure}

The observed values for the SM parameters suggest a large value for this instability scale of about $\SI{e10}{\giga\electronvolt}$ \cite{Degrassi:2012ry,Buttazzo:2013uya}. As illustrated in Fig.~\ref{fig:metastability-bound}, the properties of the Higgs potential in the infrared (IR) are controlled by the mass parameter. For a small mass parameter, the potential has a metastable minimum in the IR similar to the EW vacuum. For large mass parameters, in particular larger than the instability scale, this vacuum disappears.
These two regimes are separated by a critical value corresponding to the \textit{metastability bound}, first introduced in Ref.~\cite{Buttazzo:2013uya} (see also Refs.~\cite{Giudice:2021viw, Khoury:2021zao,Steingasser:2023ugv,Steingasser:2024hqi,Detering:2024vxs}),
\begin{equation}\label{eq:metastability-bound}
    \mu_H^2 \lesssim \mu_{\rm crit}^2 \equiv  \frac{1}{2} e^{-3/2} |\beta_\lambda (\mu_I) | \mu_I^2.
\end{equation}
Note that the left-hand side of the metastability bound is not the physical Higgs mass but the Lagrangian parameter. At tree-level, these are simply related by $m_h^2=2 \mu_H^2$. This bound is generated through dimensional transmutation from the RG evolution of the Higgs self-coupling and can thus be exponentially separated from the scale of UV completion, $\Lambda_\text{UV} \gg \mu_I$, see Refs.~\cite{Giudice:2021viw, Khoury:2021zao} for more details.

The EW hierarchy problem may then be understood through the metastability bound if a mechanism of vacuum selection preferentially selecting a minimum in the IR with broken EW symmetry generates a sufficiently tight bound \cite{Giudice:2021viw,Khoury:2019yoo,Khoury:2021zao,Benevedes:2024tdq}. The metastability bound is however agnostic to the nature of the vacuum selection mechanism, even if not independent of their existence. Viable realisations are briefly reviewed in the following section.

In contrast to conventional natural solutions to the hierarchy problem, the scale of UV completion of the SM does not have to be a loop factor larger than the Higgs mass, $m_h^2 \sim \text{loop-factor} \times \Lambda_\text{UV}^2$, but can be exponentially separated from the instability scale, $\mu_I \ll \Lambda_\text{UV}$ \cite{Khoury:2021zao}. In the SM, the large instability scale constrains the mass parameter only to be below $\mathcal{O}(\SI{e10}{\giga\electronvolt})$, leaving the remaining hierarchy unexplained. As we discuss in Sect.~\ref{sec:models}, well-motivated extensions of the SM can however exhibit an instability at lower scales around the TeV.

\section{Mechanisms achieving Higgs metastability}\label{sec:Mechs}

An important feature of the metastability bound in Eq.~\eqref{eq:metastability-bound} is that it follows immediately from the structure of the Higgs potential and is sensitive only to the running of the quartic coupling. This implies in particular that whether or not the Higgs mass is controlled by this inequality can be answered independently from the question of \textit{why} this is the case. This generality represents one of the strongest motivations to consider this scenario, in addition to its distinct experimental signatures.

Nevertheless, and for illustration, we briefly discuss in this section three examples of fundamentally different approaches justifying a metastable vacuum from underlying physics. These encompass: i) a dynamical attractor in the early Universe; ii) statistical arguments in theories with a large number of vacua; and iii) a cosmological crunch argument.

\subsection{Self-organised localisation}

Self-organised criticality of a scalar field undergoing quantum fluctuations during inflation can localise it near the critical point on the verge of a quantum phase transition. The stochastic dynamics of the scalar field are described by a probability distribution governed by the Fokker-Planck equation with a volume term---the Fokker-Planck Volume (FPV) equation---to account for Hubble patches with higher energy densities further up the potential inflating more. If the scalar field is coupled such that it effectively varies another parameter as it evolves, it may be responsible for a quantum phase transition at a critical point along its potential. The equilibrium solutions to the FPV equation in the appropriate quantum regimes are then indeed localised close to the critical point, indicating that the volume distribution that dominates the landscape after inflation ends is one that would appear to its inhabitants as curiously fine-tuned. This phenomenon of Self-Organised Localisation (SOL) in quantum cosmology, introduced in Ref.~\cite{Giudice:2021viw}, could be responsible for the metastability of the EW vacuum if the fluctuating scalar is dynamically scanning the Higgs quadratic term and triggers the EW phase transition. In this SOL scenario, the Higgs mass is predicted to be close to the upper limit set by the vacuum metastability bound. 

\subsection{Dynamical criticality on the landscape}

The metastability bound also emerges from considerations on timescales significantly exceeding the lifetime of the EW vacuum, as discussed in Refs.~\cite{Khoury:2019yoo,Khoury:2019ajl,Kartvelishvili:2020thd,Khoury:2021zao,Khoury:2022ish,Khoury:2023ktz}. These authors consider a theory with a large number of vacua, which they suggest to model as a network: each node represents a vacuum, while the tunneling between different vacua corresponds to the links between them. In this picture, an eternal observer (experiencing several vacuum decay events) can be understood as ``exploring'' the network. This allows for the definition of the \textit{accessibility measure}, which favors regions in the network permitting an efficient exploration. Using well-established results from the study of networks and several reasonable assumptions, it can then be shown that vacua in such a region have a characteristic lifetime of order of their \textit{Page time},
\begin{equation}
	\tau_{\rm ideal}\sim \tau_{\rm Page}\sim \frac{M_{\rm Pl}^2}{H_0^3} \simeq 10^{130}\ {\rm years},
\end{equation}
where we have in the last step specified the observed value of the Hubble constant, $H_0\approx 67.4 \frac{\rm km}{\rm s \cdot Mpc}$~\cite{Buttazzo:2013uya}. 

This result can be related to the hierarchy problem using the more general arguments presented in Refs.~\cite{Khoury:2021zao,Benevedes:2024tdq}. These works argue that for the cases of a negative and positive Higgs quadratic term, respectively, that metastability of the EW vacuum not only requires the bound~\eqref{eq:metastability-bound}, but also an additional hierarchy between the instability scale and the natural value of the Higgs mass.\footnote{These arguments rely on the additional assumption that the metastability is linked to the running of the quartic coupling and not achieved through higher-dimensional operators alone. In Ref.~\cite{Enguita:2025ybx} it is demonstrated how this naturally arises in a concrete model.} Remarkably, the bound for the case of a positive quadratic term is stronger than its counterpart for a negative term precisely in the regime in which these bounds are closest to the observed value of the Higgs mass.

\subsection{Radiatively generated vacua}

The EW vacuum appears not only special due to its fine-tuned Higgs mass and quartic coupling, but also its total energy appears subject to fine-tuning, relating to the cosmological constant problem. This can be understood as our vacuum having an energy density that is just barely enough to prevent it from undergoing a cosmological crunch, while the true vacuum can be expected to suffer precisely this fate. Moreover, while the existence of the EW vacuum appears highly sensitive to the precise values of the parameters in the potential, it can be expected that relatively small changes in the parameters -- corresponding to different vacua of the selection sector -- will not lead to any qualitative changes in the UV minimum. 

In Ref.~\cite{Benevedes:2025qwt}, it was pointed out that this simple observation offers a path towards a straightforward justification of the metastability bound in Eq.~\eqref{eq:metastability-bound}. Assuming that the properties of the landscape always lead to a negative energy density in the ``natural'' vacuum, the only regions of space that avoid crunching are the ones in which the parameters of the Higgs sector are such that radiative corrections can lead to the formation of an additional minimum ``on top'' of a local maximum, elevated above its natural counterpart. See Fig.~\ref{fig:metastability-bound}. Furthermore demanding a negative Higgs quadratic term, this immediately leads to the metastability bound.

Moreover, it is straightforward to see that the elevation of interest in this argument is primarily controlled by the hierarchy $\mu_I^2 \ll \Lambda^2$. Considering for simplicity the extension of the SM potential by a dimension-six term, this amounts to a combined hierarchy
\begin{align}\label{eq:totalhierarchy}
    \mu_H^2 \lesssim \frac{1}{2} e^{-3/2} |\beta_\lambda (\mu_I)| \mu_I^2 \lesssim f(\Delta V) \frac{e^{-1/2}}{48 C_6} |\beta_\lambda (\mu_I)|^2 \Lambda_{\rm UV}^2.
\end{align}
where $f(\Delta V)$ takes values between $0$ and $1$, where smaller values correspond to a higher energy density in the false vacuum. From this perspective, the smallness of the cosmological constant can be understood as requiring the least amount of fine-tuning in the ratio $\mu_I^2/\Lambda^2$.

\section{Models with a lowered metastability bound}\label{sec:models}
The high instability scale in the SM cannot be used for the purpose of our argument since it results in a large, unexplained hierarchy between the metastability bound and the observed value of the Higgs mass. The expectation from Higgs metastability (as a solution to the EW hierarchy problem) is then new physics capable of tightening the metastability bound, thus removing the unexplained scale separation in the SM: 
Eq.~\eqref{eq:metastability-bound} implies that the metastability bound on $m_h$ is only separated from the instability scale $\mu_I$ by a factor of $\beta_\lambda \simeq \mathcal{O}(10^{-1})$. Improving the bound to fully justify the observed $m_h$ as a by-product of metastability requires then that $\mu_I \sim \mathcal{O}(\text{TeV})$. 

In this context, it is important to recall that the inequality~\eqref{eq:metastability-bound} is to be understood as a strict upper bound. Its saturation corresponds to a potential with no potential barrier protecting the EW vacuum, suggesting that stability against fluctuations would require an additional hierarchy. The bound can further be strengthened by taking into account that for a nearly saturated bound, the EW vacuum is given by a near-inflection point, corresponding to an additional suppression of the physical mass of up to $1/e$~\cite{Benevedes:2025qwt}. Throughout this section, we will neglect these additional factors and only report the most conservative estimate assuming the tree-level relation $m_h^2=2 \mu_H^2$ and Eq.~\eqref{eq:metastability-bound}.

In the following, we briefly review two examples for models naturally lowering the instability scale while upholding the metastability of the EW vacuum, as well as an in this sense incomplete model demonstrating to which extent the instability scale can be lowered.

\subsection{Axion-Higgs criticality}
Light new physics in the form of an extended scalar sector is capable of effectively lowering the instability scale. A well-motivated candidate is an axion-like particle (ALP), which can be naturally light as a pseudo-Goldstone boson.
Consider an ALP $a$ mixing with the Higgs, described by the low-energy scalar potential \cite{Detering:2024vxs,Harigaya:2023bmp},
\begin{equation}
    V(H,a) = - \mu_H^2 H^\dagger H + \lambda_H (H^\dagger H)^2 + \frac{1}{2} m_a^2 a^2 - \frac{1}{2} A \sin{(\delta)} a \left( H^\dagger H - \frac{v^2}{2}\right),
\end{equation}
where $v^2 = 2 \left<H^\dagger H\right>$ is the vacuum expectation value of the Higgs, $A$ is the dimensionful coupling between the ALP and the Higgs, and $\delta$ is a CP violating angle that can arise in the UV completion of this theory \cite{Detering:2024vxs}. This SM extension has two implications for the metastability bound. Firstly, matching the Axion-Higgs model to observations reduces the quartic coupling at the EW scale with respect to the SM. Secondly, the metastability bound Eq.~\eqref{eq:metastability-bound} is modified in the two-scalar potential \cite{Detering:2024vxs},
\begin{equation}
    \mu_H^2 \lesssim \mu^2_\text{crit} = - \frac{1}{2} e^{-3/2} \beta_\lambda(\Tilde{\mu}_I)  \Tilde{\mu}_I^2 + \frac{1}{2} \frac{v^2 A^2 \sin^2{\delta}}{m_a^2},
\end{equation}
where now $\Tilde{\mu}_I$ is the renormalisation scale at which the effective quartic coupling $\Tilde{\lambda}_\text{eff}$ along the flat direction in $a$ vanishes \cite{Detering:2024vxs},
\begin{equation}
    \Tilde{\lambda}_\text{eff} \equiv \lambda_{H} - \frac{1}{2} \frac{A^2 \sin^2{\delta}}{m_a^2}.
\end{equation}
Similar to the SM Higgs sector, the Axion-Higgs model does not exhibit an IR vacuum for too large Higgs mass parameters. Requiring the existence of the IR vacuum sets a tighter metastability bound on the Higgs mass parameter than the SM, rendering the Higgs mass near-critical in parts of the parameter space.
\begin{figure}[t]
    \centering
    \includegraphics[width=\linewidth]{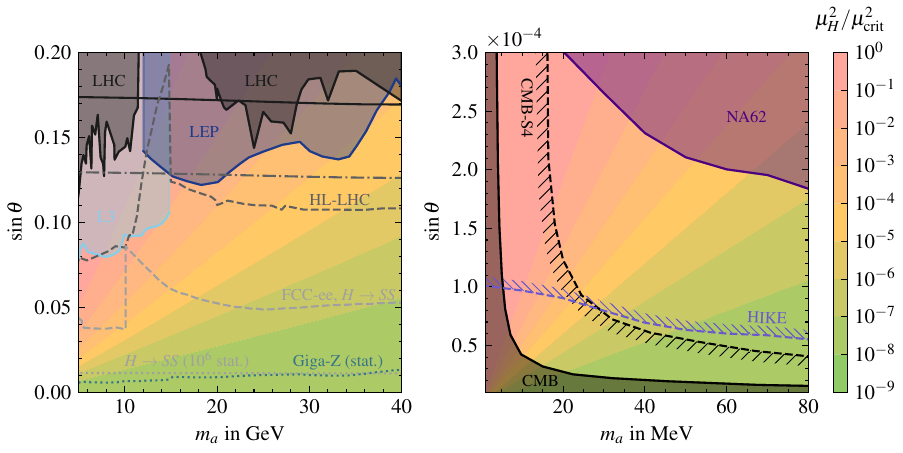}
    \caption{\it Critical value of the Higgs mass parameter in the Axion-Higgs model. The contours of the remaining hierarchy between the observed Higgs mass parameter in the Axion-Higgs model and the metastability bound are shown from red (small hierarchy) to green (large hierarchy). Existing constraints on the parameter space are shaded and projected experimental sensitivities are indicated by dashed and dotted lines. \cite{Detering:2024vxs} }
    \label{fig:alp-metastability-bound}
\end{figure}
The metastability bound in the Axion-Higgs model as a function of the two new model parameters chosen to be the new scalar mass and its mixing angle with the Higgs is shown in Fig.~\ref{fig:alp-metastability-bound}. The axion parameter space motivated from Higgs metastability in the Axion-Higgs model turns out to feature an ALP with masses on the \si{MeV} or \si{GeV} scale. This parameter space is partially constrained by collider searches and cosmological bounds, and can be further probed by upcoming and proposed experiments.

Prominent phenomenological signatures of an ALP in \si{\giga\electronvolt} range mixing with the Higgs are sizeable exotic Higgs branching ratios, and direct production at $Z$ factories through the mixing-induced $SZZ$ vertex \cite{Detering:2024vxs}. Direct production (dashed) \cite{L3:1996ome,LEPWorkingGroupforHiggsbosonsearches:2003ing,OPAL:2002ifx,Fuchs:2020cmm} and exotic Higgs decays (dash-dotted) \cite{Carena:2022yvx,ATLAS:2021vrm,deBlas:2019rxi} at HL-LHC are expected to further probe the motivated parameter space in the \si{\giga\electronvolt} range \cite{deBlas:2019rxi}. The entire motivated parameter space for \si{\giga\electronvolt} masses can be comprehensively tested at the FCC-ee \cite{Carena:2022yvx} and future Z factories \cite{FCC:2018evy,Bernardi:2022hny}.

ALPs with masses in the \si{MeV} range can be probed through rare meson decays. Future flavour factories \cite{HIKE:2022qra} can greatly increase the sensitivity in the interesting parameter space \cite{Antel:2023hkf,Beacham:2019nyx}. Complementary to collider searches are constraints derived from bounds on the effective number of relativistic degrees of freedom from CMB data \cite{Ibe:2021fed}. Future observatories such as CMB-S4 \cite{CMB-S4:2016ple} as a benchmark provide a constructive complementarity to virtually test all parameter space in the MeV range.

In conclusion, Axion-Higgs criticality predicts an axion in the MeV or GeV range, and importantly, virtually all motivated parameter space can be explored by upcoming and proposed experiments.

\subsection{Majoron model}\label{sec:Majoron}

In a large class of viable BSM scenarios, a strict metastability bound on $m_h$ can be achieved through a combination of
\begin{itemize}
    \item[i)] Heavy exotic fermions, which lower the value of $\mu_I$ via their renormalisation group (RG) impact on the running of $\lambda$. 
    Taken by themselves, they may further destabilise the vacuum, though.
    \item[ii)] At least one BSM scalar to stabilise the vacuum and ensure a lifetime of the EW larger than the age of the Universe, by stabilising the vacuum at some scale above $\mu_I$.
\end{itemize}
\begin{wrapfigure}{l}{0.45\textwidth}
    \centering
    \includegraphics[width=\linewidth]{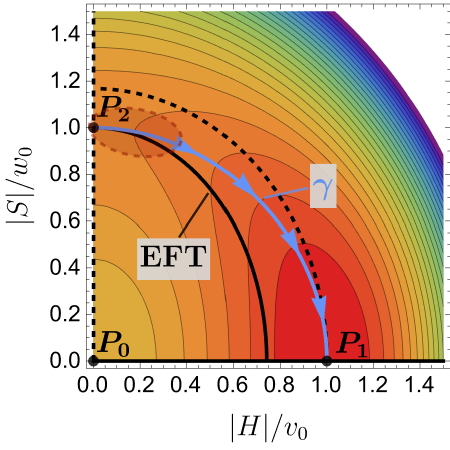}
    \caption{\it Favoured vacuum configuration, normalized as in Ref.~\cite{Enguita:2025ybx}.}  
    \label{fig:tunneling:potential+gradient-descent}
\end{wrapfigure}
In the context of the Higgs mass metastability bound, type-I seesaw fermions were considered in Refs.~\cite{Steingasser:2023ugv, Khoury:2021zao,Benevedes:2024tdq}, while the effect of scalar physics has been explored via a mass-dimension six ($d=6$) scalar operator~\cite{Steingasser:2023ugv}, in the spirit of an EFT. In the regime corresponding to a tight metastability bound, the lifetime of the EW vacuum becomes very dependent on the tower of higher-dimensional scalar operators, suggesting the need to use complete ultraviolet theories for a proper treatment.

The Majoron model of neutrino masses~\cite{Chikashige:1980ui} is a complete ultraviolet model, which naturally contains both ingredients enumerated above: heavy neutral leptons (HNLs) as in the type-I seesaw, plus a scalar field whose vacuum expectation value (VEV) sets the overall scale of both the mass $M_N$ of the heavy Majorana neutrinos  and   the mass $M_s$ of the radial scalar field $S$. This fact automatically addresses the subtle question of the proximity of $\mu_I$ and the new physics scale to obtain a tight bound. To be precise, it encompasses the condition
\begin{equation}
M_N \lesssim \mu_I \lesssim M_s\,,
\label{scales}
\end{equation}
with all the scales naturally close within a few orders of magnitude.

To have $M_N, \mu_I \sim \mathcal{O}(\text{TeV})$ while simultaneously generating tiny masses for the observed neutrinos is possible in
the large class of so-called ``low-scale'' Majoron models, which exhibit an approximate $U(1)$ lepton-number symmetry in different realisations~\cite{Branco:1988ex,Kersten:2007vk,Abada:2007ux,Moffat:2017feq, Mohapatra:1986aw,Mohapatra:1986bd,Akhmedov:1995ip,Malinsky:2005bi,Shaposhnikov:2006nn}. 
In the context of the metastability bound on $m_h$, the low-scale Majoron paradigm has been recently explored in Ref.~\cite{Enguita:2025ybx}, showing that the hierarchy of scales in Eq.~(\ref{scales}) can naturally be encompassed around the TeV range and leads to strong $m_h$ bounds.  

The two-field  tree-level potential reads, denoting by $\kappa$ the $H$-$S$ portal,  
\begin{align}
    &V(H, S)\supset-\mu_H^2 H^\dagger H -\mu_S^2 |S|^2+
\lambda_H (H^\dagger H)^2 +\lambda_S |S|^4+\kappa\,  H^\dagger H |S|^2\,,%+\, V_{\slashed{L}}   \,,
\label{eq:majoron:potential}
\end{align}
where at low-energies, and for non-trivial VEV of S, the tree-level effective Higgs self-coupling is $\lambda_\text{eff}\equiv \lambda_H -{\kappa^2}/{4\lambda_S}$. The analysis of its phases in Ref.~\cite{Enguita:2025ybx} identifies a unique suitable configuration illustrated in \cref{fig:tunneling:potential+gradient-descent}. Radiative corrections generate a barrier around the tree-level saddle point $\bld{P_2}$, which then gives rise to the EW--metastable--vacuum. Tunnelling towards the true minimum in $\bld{P_1}$ takes place close to the blue field-space trajectory $\gamma$ parametrised as a function of the two fields $S$ and $H$~\cite{Enguita:2025ybx}. Its significant deviation from the curve representing the naive EFT around the EW vacuum in black emphasises the importance of working with a complete model. 

\begin{figure}[t]
    \centering
    \begin{subfigure}[t]{0.46\textwidth}
    \vspace*{-7.2cm}
    \includegraphics[width=\textwidth]{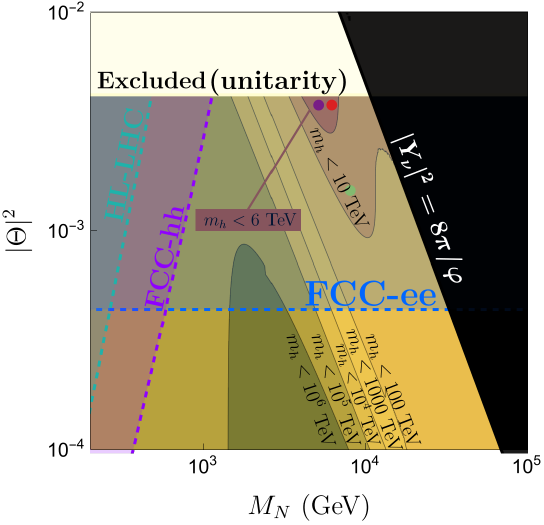}{}
    \end{subfigure}
    \begin{subfigure}[t]{0.47\textwidth}
    \includegraphics[width=\textwidth]{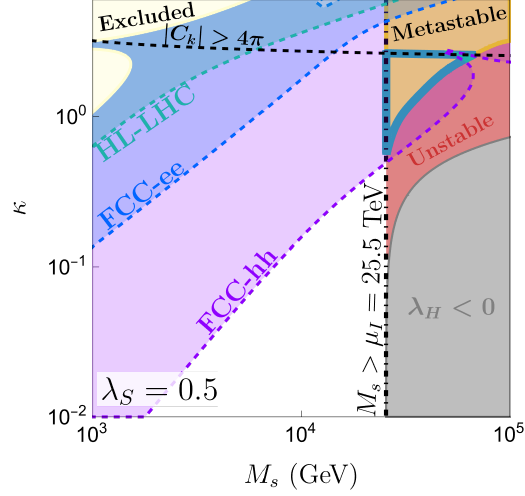}{}
    \end{subfigure}
    \vspace*{-0.43cm}
    \caption{\it \textbf{Left:} HNL space, showing the overlap of future collider sensitivity regions with  metastability bounds. \textbf{Right:} Scalar space for the HNL benchmark red point of the left panel and $\lambda_S=0.5$. The triangular-like area shows the FCC-hh metastability reach.}
    \label{fig:majoron-results}
\end{figure}

The lowering of the metastability bound depends mainly on the fermionic HNL sector, as depicted on the left panel of \cref{fig:majoron-results} as a function of the set $\{M_N, |\Theta|^2\}$ (where $\Theta$ denotes the trace of the light-heavy neutrino mixing matrix~\cite{Enguita:2025ybx}). There, shaded contours correspond to different metastability bounds on $m_h$, while the excluded regions are dominated by PMNS unitarity constraints and perturbativity. The scalar sector is instead crucial to the lifetime of the EW vacuum, which allows only the golden region of the $\{M_s, \kappa\}$ shown in the right panel of \cref{fig:majoron-results} for the HNL benchmark point in red on the left and $\lambda_S=0.5$. Similar results hold for other HNL and $\lambda_S$ values~\cite{Enguita:2025ybx}.

As a stringent metastability bound brings the framework into the TeV range,  the question of its testability at present and future colliders is peremptory. The left panel of \cref{fig:majoron-results} also demonstrates the {\it solid prospects of FCC-ee to scan the entire HNL allowed region of strong metastability constraints}, up to Higgs mass bounds of a few tens of TeV (in contrast to the (very) modest reach of the FCC-hh /HL-LHC).

The signals stemming from the scalar sector could instead be tackled at the FCC-hh, as illustrated in the right panel of \cref{fig:majoron-results} for the chosen benchmark point. {\it The allowed region of interest which meets all requirements is the triangular-like area in \cref{fig:majoron-results}, delimitated by a thick continuous  till line}. Its vertical boundary corresponds to $M_s \ge \mu_I$; the roughly horizontal boundary ensures perturbativity; the non-trivial boundary to the right is due to the interplay of two effects: larger  $\kappa$ values generically stabilise the EW vacuum, and increasing $M_s$ weakens the stabilising impact of the heavy scalar. Similar results hold for large sets of values of the fermionic and scalar parameters of the theory, see Ref.~\cite{Enguita:2025ybx}.

In summary, a nice complementarity appears between the FCC-ee and the FCC-hh as to their ability to test Higgs criticality, with FCC-ee tackling large regions of the fermion parameter space of interest and FCC-hh part of the corresponding scalar domain. For the most stringent bounds analysed (i.e., $m_h<6$ TeV, and even for $m_h< 10$ TeV), there is a good chance that BSM signals appear both at the FCC-ee {\it and} the FCC-hh.

\subsection{Singlet-doublet model}
Mechanisms relying on fermions to lower the instability scale are limited by compliance with existing observations. One particularly interesting model from this perspective is the singlet-doublet model~\cite{Mahbubani:2005pt,DEramo:2007anh,Enberg:2007rp,Cohen:2011ec,Cheung:2013dua,Abe:2014gua,Calibbi:2015nha,Freitas:2015hsa,Banerjee:2016hsk,Cai:2016sjz,LopezHonorez:2017zrd,Arcadi:2019lka,Fraser:2020dpy,Homiller:2024uxg}, whose phenomenology with a special focus on the metastability bound has recently been studied in Ref.~\cite{Benevedes:2025qwt}. While this model in itself is ``incomplete'' in the sense that additional new physics is necessary to avoid an unstable vacuum, it nevertheless serves as an important proof of concept to demonstrate the extent to which new physics can lower the instability scale in compliance with existing bounds.

This model involves one singlet, $\psi_L$, and a pair of $SU(2)$ doublets $\chi_{L,R}$ with hypercharge $1/2$, whose total Lagrangian is given by
\begin{align}
    \mathcal{L}_{\psi\chi} &= i \left( \overline{\psi_L} \slashed{\partial} \psi_L + \overline{\chi_L} \slashed{D} \chi_L  + \overline{\chi_R} \slashed{D} \chi_R \right) - \left(\frac{1}{2}m_S \psi_L \psi_L + m_D \overline{\chi_R} \chi_L + {\rm ~h.c.} \right) \nonumber \\
    &-y_1 \chi_L \tilde{H} \psi_L - y_2 \overline{\chi_R} H \psi_L - \lambda_i L_L^i H \psi_L - \lambda'_i \chi_L H \overline{e_R}^i + {\rm ~h.c.} \label{eq:Linos}
\end{align}
These new fermions are analogous to the Higgsino and bino arising from the Minimal Supersymmetric Standard Model (MSSM) in the limit of a decoupled wino. In order to keep results manageable, the authors of Ref.~\cite{Benevedes:2025qwt} focused on the special case $\lambda_i=\lambda_i^\prime =0$, corresponding to an additional $\mathbb{Z}_2$ symmetry for the BSM fermions. 

\begin{figure}[h]
\centering
    \includegraphics[width=0.45\linewidth]{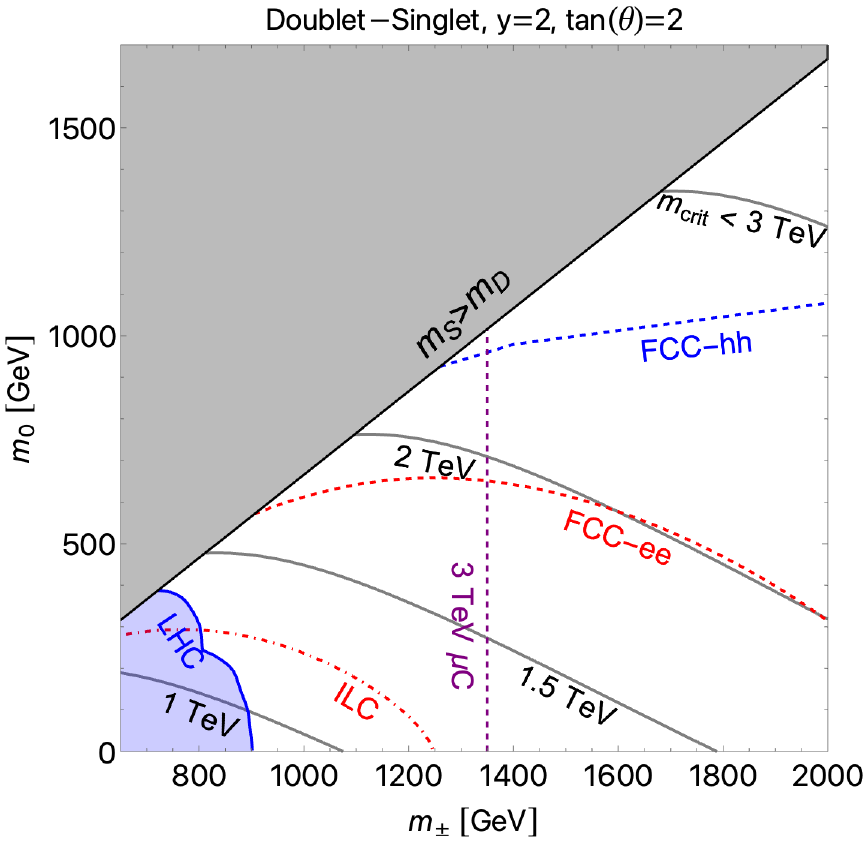}
    \includegraphics[width=0.45\linewidth]{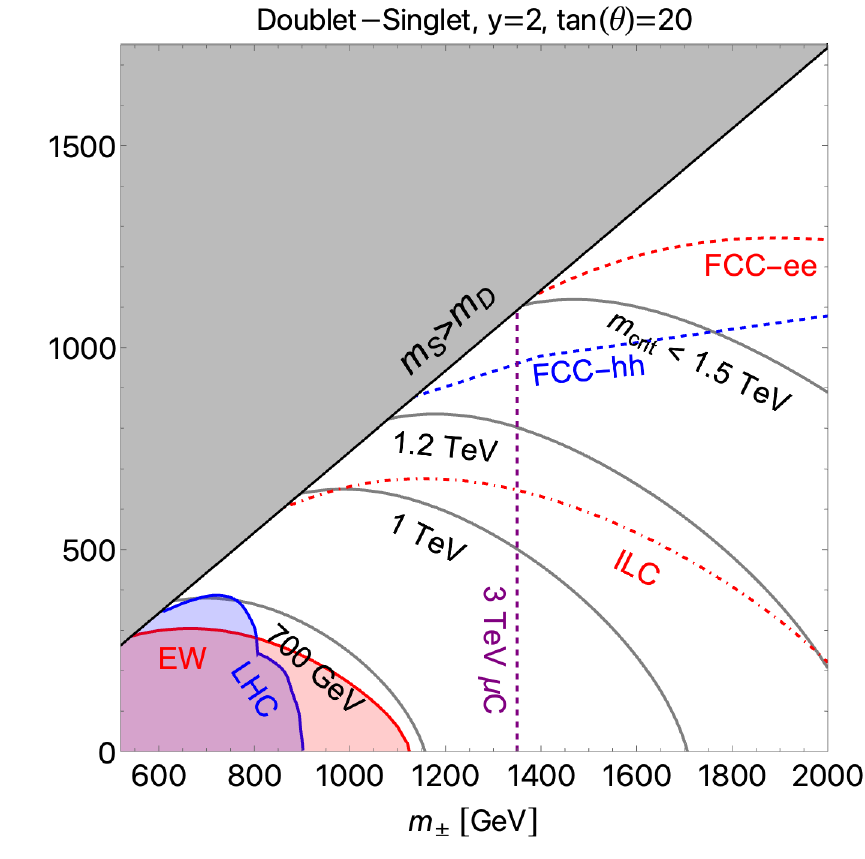}
    \caption{\it The metastability bound as a function of the fermion masses $m_D$ and $m_0$ for $y=2$ and two benchmark values for $\theta$, assuming $m_S<m_D$. The red solid line and shaded region (in the right panel) represent existing experimental constraints from EW precision observables~\cite{Baak:2014ora}, and the red broken lines in both panels are projected reach from EW precision at FCC-ee (dashed)~\cite{Fan:2014vta,deBlas:2016nqo} and ILC (dash-dotted)~\cite{Fan:2014vta}. The solid blue line and shaded region shows present-day constraints from direct searches at the LHC~\cite{ATLAS:2019lff,ATLAS:2021yqv,ATLAS:2022zwa,CMS:2022sfi,ATLAS:2022hbt,CMS:2024gyw}, and the dashed blue line shows projected reach from direct searches at FCC-hh~\cite{Mangano:2017tke}. The broken purple vertical lines show the projected reach of direct searches for charged particles production at a future $3 \, \text{TeV}$ muon collider~\cite{Homiller:2024uxg}.}
    \label{fig:inos}
\end{figure}

The results of this work are shown in Fig.~\ref{fig:inos}. For this figure, the Yukawa couplings are expressed through the effective parameters $y_1 = y \cos \theta$ and $y_2 = y \sin \theta$, as well as as the physical masses $m_{\pm}$ and $m_0$: the Lagrangian~\eqref{eq:Linos} leads to a non-diagonal mass matrix. In the limit $y_1 v, y_2 v\ll m_D,m_S$, the five fermionic degrees of freedom in this theory combine into an electrically charged Dirac fermion with mass $m_\pm \simeq m_D$ and three neutral Majorana fermions. The lightest of these has a mass $m_0\simeq m_S/3+2 (3 \sqrt{3})^{1/3} \sqrt{m_D^2 + m_S^2/3}\cos[(\theta+4\pi)/3]$. 

We present the results of Ref.~\cite{Benevedes:2025qwt} in Fig.~\ref{fig:inos}. Remarkably, these authors find that this model allows for a metastability as strong as $\mathcal{O}$(few hundred)~GeV. Just as for the HNLs considered in Sec.~\ref{sec:Majoron}, the parts of parameter space corresponding to the most stringent bounds lie within reach of future lepton colliders. 

\newpage
\section{Conclusion}

The hierarchy problem poses a serious challenge to the paradigm of naturalness, which has served as one of the guiding principles of fundamental physics model building for decades. In this report, we have reviewed the first steps in the development of an alternative approach to explaining the Higgs mass while upholding the biggest advantage of naturalness --- its predictability, manifesting through a mostly model-independent scale of new physics through the so-called metastability bound.

The metastability bound arises solely from the properties of the Higgs' effective potential and the running of its quartic coupling. Its existence and possible application to the hierarchy problem does not commit to any particular BSM physics so long as the vacuum instability scale is lowered and a vacuum selection mechanism preferentially selects a near-critical Higgs in the landscape. And indeed, in Sec.~\ref{sec:Mechs} we have reviewed three entirely independent vacuum selection mechanisms, all of which lead to this particular bound. It is likely that further mechanisms can and will be brought forward.

While the testability of the vacuum selection mechanism depends on how it is realised, the metastability bound itself is independent of this and is falsifiable by virtue of its connection to the instability scale. If it plays a role in explaining the Higgs mass, the instability scale in the SM of order $\mathcal{O}(10^{10})$~GeV must be lowered closer to the TeV scale to be a sensible upper bound on the observed Higgs mass value of $125$~GeV. These mechanisms therefore require new physics capable of lowering the instability scale without destabilising the EW vacuum entirely, in addition to the new physics responsible for the vacuum selection. 

In Sec.~\ref{sec:models}, we have reviewed two examples for SM extensions in which a lowered vacuum instability scale is naturally realised. An axion-like particle with mass $\sim\mathcal{O}$(MeV-GeV) can lower the instability scale through a threshold correction while ensuring a long enough vacuum lifetime; the axion does not introduce any new marginal couplings, therefore the beta functions remain unchanged. In the Majoron model, the instability scale is lowered by the HNLs' contribution to the running of the quartic coupling, while the additional scalar responsible for the HNLs' mass also ensures metastability of the EW vacuum. 

In both of these realisations the regions of natural parameter space allowing for a strong bound lie within range of proposed experiments such as FCC. The entire motivated parameter space for the ALP with a GeV scale mass can be probed by direct searches and exotic Higgs decays at the FCC-ee and a Giga-Z factory. The complementarity of proposed (benchmark) CMB measurements and flavour factories can comprehensively test near-criticality for MeV scale ALPs. In the Majoron model, a significant lowering of the instability scale requires a combination of large Yukawa couplings and relatively light HNLs, which immediately translates to a strong signal in FCC-ee. Similarly, the scalar stabilises the vacuum effectively only for a large portal coupling and a mass only slightly above the instability scale, corresponding to a strong signal in FCC-hh. 

These examples demonstrate the potential impact that the metastability bound can have on model building, phenomenology, and novel experimental signatures related to the hierarchy problem. Assuming the existence of \textit{some} mechanism favouring a metastable vacuum (or directly imposing Eq.~\eqref{eq:metastability-bound}), the hierarchy problem can be addressed in any model capable of significantly lowering the instability scale without destabilising the vacuum too much. While a definite solution to the hierarchy problem still requires identifying this vacuum selection mechanism, this task is effectively decoupled from model building near the EW scale, with the latter being in range of realistic experiments.

\newpage
\bibliographystyle{JHEP}
\bibliography{references}

\end{document}